# Tungsten isotope evolution during Earth's formation and new constraints on the viability of accretion simulations

D.C. Rubie[1*], K.I. Dale[2], G. Nathan[3], M. Nakajima[4], E.S. Jennings[5], G.J. Golabek[1], S.A. Jacobson[3], A. Morbidelli[6,2]

[1] Bayerisches Geoinstitut, University of Bayreuth, Universitätsstrasse 30, 95440 Bayreuth, Germany

[2] Université Côte d'Azur, Observatoire de la Côte d'Azur, CNRS, Laboratoire Lagrange, Nice, France

[3] Department of Earth & Environmental Sciences, Michigan State University, 288 Farm Lane, East Lansing, MI 48823, USA

[4] Department of Earth and Environmental Sciences, University of Rochester, 227 Hutchison Hall, Rochester, NY 14627, USA

[5] School of Natural Sciences, Birkbeck, University of London, Malet Street, London WC1E 7HX, UK

[6] Collège de France, CNRS, PSL Univ., Sorbonne Univ., Paris, 75014, France

* Corresponding author:
  E-mail address: dave.rubie@uni-bayreuth.de

**The Hf-W isotopic system is the reference chronometer for determining the chronology of Earth's accretion and differentiation. However, its results depend strongly on uncertain parameters, including the extent of metal-silicate equilibration and the siderophility of tungsten. Here we show that a multistage core-formation model based on N-body accretion simulations, element mass balance and metal-silicate partitioning, largely eliminates these uncertainties. We modified the original model of Rubie et al. (2015) by including (1) smoothed particle hydrodynamics estimates of the depth of melting caused by giant impacts and (2) the isotopic evolution of $^{182}$W. We applied two metal-silicate fractionation mechanisms: one when the metal delivered by the cores of large impactors equilibrates with only a small**

**fraction of the impact-induced magma pond and the other when metal delivered by small impactors emulsifies in global magma oceans before undergoing progressive segregation. The latter is crucial for fitting the W abundance and $^{182}$W anomaly of Earth's mantle. In addition, we show, for the first time, that the duration of magma ocean solidification has a major effect on Earth's tungsten isotope anomaly. We re-evaluate the six Grand Tack N-body simulations of Rubie et al. (2015). Only one reproduces $\varepsilon^{182}$W=1.9±0.1 of Earth's mantle, otherwise accretion is either too fast or too slow. Depending on the characteristics of the giant impacts, results predict that the Moon formed either 143-183 Myr or 53-62 Myr after the start of the solar system. Thus, independent evaluations of the Moon's age provide an additional constraint on the validity of accretion simulations.**

## 1. Introduction

The most important constraint on the timing of Earth's accretion and core formation is the Hf-W isotopic system. $^{182}$Hf decays to $^{182}$W with a half-life of 8.9 Myr and, because Hf is a lithophile element and W is moderately siderophile, early planetary core formation results in an excess of $^{182}$W in the mantle relative to undifferentiated chondritic concentrations (Jacobsen, 2005; Kleine et al., 2004, 2009; Rudge et al., 2010). Earth's mantle has a positive $^{182}$W anomaly and over the past twenty years attempts have been made to elucidate its cause, especially in terms of the chronology of core formation. Early studies either modeled accretion and differentiation as a continuous process or considered core formation to be a single-stage event, neither of which is realistic. More recent studies have used the results of astrophysical N-body accretion simulations as the basis for interpreting the $^{182}$W anomaly of Earth's mantle (Nimmo and Agnor, 2006; Nimmo et al., 2010; Fischer and Nimmo, 2018; Zube et al., 2019). All studies have concluded that several important factors contribute to the $^{182}$W anomaly, in addition to the timing of accretion and core formation. These include the fraction of accreted metal that equilibrates with silicate mantle, the fraction of silicate mantle that equilibrates with accreted metal, and the metal-silicate partition coefficient of tungsten which is a strong function of oxygen fugacity (Jennings et al., 2021). In fact, some studies (e.g., Rudge et al., 2010) have shown that such factors may be just as important in determining the W isotope anomaly of Earth's mantle as the chronology of accretion and differentiation. In addition, discussions of the fractions of equilibrating metal and silicate have seldom considered plausible physicochemical constraints on these variables. Uncertainties in such factors have therefore precluded the use of the Hf-W system to reliably discriminate between viable and non-viable models of Earth's accretion (Zube et al., 2019).

Recent studies have made considerable progress in combining the output of planetary N-body accretion simulations with models of multi-stage core formation (Rubie et al., 2015, 2016; Fischer et al., 2017; Blanchard et al., 2022; Dale et al., 2023; Gu et al., 2023). The aim is to refine the pressures of metal-silicate equilibration and the chemistry of the accreting bodies as a function of their heliocentric distances of origin by fitting the results to the elemental chemistry of Earth's primitive mantle (i.e., the bulk silicate earth composition, BSE). Using a rigorous mass balance/element partitioning approach (Rubie et al., 2011), the calculated mantle concentrations of 17 elements: Mg, Fe, Si, Ni, Co, Nb, Ta, V, Cr, S, Pt, Pd, Ru, Ir, W, Mo, and C have been fit to BSE concentrations, mostly with a high degree of success (Rubie et al., 2015, 2016; Blanchard et al., 2022; Jennings et al, 2021). Only the calculated mantle concentrations of W and Mo are too high, by up to a factor of 3-4 in the case of W.

Here we present a new metal-silicate fractionation model which is applicable when the metal from accreted objects becomes dispersed in a turbulently-convecting magma ocean as small cm-size liquid metal droplets (Fig. 1b). By incorporating this model into modified accretion/core formation simulations of Rubie et al. (2015), we eliminate the problem of large excesses of W in the mantle determined by Jennings et al. (2021) which allows us to address the evolution of the radiogenic $^{182}$W isotopic anomaly of Earth's mantle.

A great advantage of this approach for modeling the tungsten isotope anomaly is that fractions of equilibrating mantle can be determined using physically-plausible hydrodynamic models (Fig. 1; Rubie et al., 2015; Deguen et al., 2011, 2014) and the average fraction of accreted metal that equilibrates can be constrained by least-squares fitting (see below). In addition to pressure and temperature, oxygen fugacities at which metal and silicate equilibrate are determined by mass balance (Rubie et al., 2011) which means that the tungsten metal-silicate partition coefficient is determined for every metal-silicate equilibration (core formation) event in the N-body simulations.

**2. Methods**

*2.1 Accretion/core formation model.* Details of our accretion/core formation model are presented by Rubie et al. (2015) and only essential details, together with modifications, are provided here. The model is based on astrophysical N-body accretion simulations that simulate the growth of the terrestrial planets in the inner solar system from a suite of accretional collisions of planetesimals and planetary embryos[1] (see Raymond et al., 2014, for a review). Here, we use Grand Tack N-body accretion simulations in which Jupiter and Saturn migrate first inwards and then back out in order to achieve a realistic mass for Mars-like bodies (Walsh et al., 2011). Future work will examine the case where the terrestrial planets grow from a narrow and dense ring of embryos and planetesimals located at around 1 au (Hansen, 2009; Nesvorný et al., 2021; Woo et al., 2024). Core-mantle differentiation during planet's growth is modeled with the aim of producing a final mantle composition that is identical to that of the bulk silicate Earth.

---

[1] Planetesimals are asteroid-like bodies, typically of a size of ~100km, which may or may not be differentiated depending on their formation time. Planetary embryos are objects with few times the mass of the Moon, possibly up to the mass of Mars, which formed from planetesimal collisions or pebble accretion during the gas-dominated phase of the protoplanetary disk. They are all considered to be differentiated.

The reader should keep in mind that the present study is largely methodological and its goal is to demonstrate how core-mantle differentiation and the chronology of Earth's accretion can be tested using any type of dynamical accretion simulation.

The naming convention used for the simulations conveys information about the starting disk parameters (Table 1). For example, in 4:1-0.5-8, "4:1" indicates the total starting mass of embryos ratioed to that of planetesimals, the second parameter indicates the initial mass of the largest embryo (0.025, 0.05 or 0.08 $M_e$) and "8" is the run number. The prefix "i" indicates that the initial mass of embryos increases with heliocentric distance in the range listed in Table 1. For simplicity, below we use the terminology "simulation 1-6" (Table 1).

The Grand Tack simulations start with around 100-200 embryos that are distributed over a heliocentric distance 0.7-3.0 au. Within the disk, several thousand planetesimals are distributed from 0.7 to 9 au. For computational reasons, the mass of each planetesimal is on the order of $10^{-4} M_e$ and, thus, they are actually tracers for a large number of much smaller bodies. Details of starting configurations are given in Table 1 and Rubie et al. (2015, Table 1).

The bulk compositions of all starting bodies are required as a function of heliocentric distance. All non-volatile elements, except oxygen and refractory elements, are present in relative CI concentration ratios. Oxygen contents are varied to define an oxidation gradient such that bodies originating at less than ~1 au are highly reduced and beyond ~1 au become increasingly oxidized with heliocentric distance due to the effects of water (Rubie et al., 2015; Monteux et al., 2018). The details of this gradient are refined by least squares minimization (Fig. 6 in Rubie et al., 2015; Fig. 8 in Jennings et al., 2021). In these previous studies the refractory element concentrations were enhanced relative to CI composition in all bodies by 22% to achieve the primitive mantle refractory element abundances (Palme and O'Neill, 2014; Rubie et al., 2011). Here a refractory element enhancement of 30% is applied, but only to the highly reduced bodies originating within ~1 au. The reason is that refractory enriched bodies should have formed only inwards of the silicate sublimation line (Morbidelli et al., 2020).

Earth was initially one of the original planetary embryos and grows through collisions with planetesimals and other embryos. Embryo collisions are highly energetic giant impacts and cause large-scale melting of the target and thus the development of deep magma ponds (Fig. 1a), each of which then hydrostatically relaxes into a global magma ocean that overlies the solid mantle. Planetesimals in contrast do not generate large scale melting because of their small mass; they either impact an already existing magma ocean, or deposit their material in the uppermost solid mantle or crust if the previous magma ocean has already crystallized. In the latter case, metal-silicate equilibration occurs only when a new global magma ocean is produced, as a consequence of the next giant impact (Fig. 1b).

In most of our previous studies, the pressures of metal-silicate equilibration, at the base the magma ponds generated by giant impacts, were estimated by least squares fitting (Rubie et al., 2015). Here, we determine the pressure at the base of magma ponds, where metal-silicate equilibration occurs, using the results of smoothed particle hydrodynamics (SPH) simulations of the extent of giant-impact induced melting (Nakajima et al., 2021; Dale et al., 2023). Parameters required for this determination (impact angle, impact velocity and the masses of the impactor and target) are output by the N-body simulations (e.g., Supplementary Material). Nakajima et al. (2021) provides results for impact angles of 0°, 30°, 45°, 60°, and 90° and for each impact we use the results for the closest of these angles (see also Gu et al., 2023). Results are presented by Nakajima et al. (2021) for target bodies with two different initial temperature distributions that have corresponding target surface temperatures of 300 K and 2000 K respectively. We primarily use the high-temperature distribution but also show the effects of using the low-temperature results. Temperatures at the base of magma ponds/oceans are assumed to lie midway between the peridotite solidus and liquidus (Rubie et al., 2015).

For the pressure of equilibration of the planetesimal material, Dale et al. (2023) assumed that the entire volume of silicate melt that is created by a giant impact evolves into a global magma ocean without any cooling and crystallization. This allows the depth of the magma ocean and hence its pressure to be estimated, again using the SPH results of Nakajima et al. (2021). However, this approach is only valid immediately following a giant impact. The magma ocean will progressively crystallize from its bottom upwards and, therefore, on average, the equilibration of planetesimal material will occur at a lower pressure. For this reason, we determine the average fraction of the core-mantle boundary (CMB) pressure at which planetesimals equilibrate in a magma ocean (one value for all impacts, for simplicity) by fitting the results to the composition of the BSE.

Regardless of whether the material comes from an embryo or a planetesimal and equilibrates in a magma pond or global magma ocean, the compositions of metal and silicate after equilibration are determined by a combination of mass balance and element partitioning that is used to determine the concentrations of major elements through the coefficients $a$, $b$, $c$, $d$, $x$, $y$ and $z$ in the following equilibrated stoichiometric compositions (Rubie et al., 2011):

$$[(FeO)_x (NiO)_y (SiO_2)_z (Mg_u Al_m Ca_n)O] + [Fe_a Ni_b Si_c O_d]$$

*silicate liquid*                                         *metal liquid*

The coefficients $u$, $m$ and $n$ are constants because they describe concentrations of lithophile elements. Concentrations of trace elements are determined by metal-silicate partitioning alone. In the present study, the partition coefficient of W is of particular importance. In addition to *P* and *T*, its partitioning is strongly affected by oxygen fugacity because W dissolves in silicate liquid with a high valence of 6+ (O'Neill et al., 2008; Jennings et al., 2021). It is also affected significantly by the metal composition, especially its carbon concentration. The latter is determined following Hirschmann et al. (2021) and Blanchard et al. (2022) with the result that that almost all carbon is accreted to Earth by fully-oxidized carbonaceous chondrite planetesimals from the outer solar system which have a bulk C concentration of 3.35 wt%. In contrast, the bulk C contents of embryos and planetesimals from the inner solar system lie in the range 0.002-0.16 wt% due to carbon loss caused by high temperatures, especially during core-mantle differentiation (Hirschmann et al., 2021; Blanchard et al., 2022). For each core-forming event we determined the metal-silicate partition coefficient of W ($D_W^{m/s}$) using the parameterizations of Jennings et al. (2021) combined with mass balance constraints.

*2.2 Metal-silicate fractionation models*

*2.2.1 Focused metal-silicate fractionation mechanism.* When a planetary embryo generates a deep magma pool or a large differentiated planetesimal impacts a global magma ocean (Fig. 1a), the volume of silicate liquid that equilibrates with the metal from the projectile's core is calculated from the radius of the impactor's core and the depth of the magma pool or magma ocean using the relationship from the hydrodynamic model of Deguen et al. (2011, 2014):

$$\emptyset_{met} = \left(\frac{r_0}{r}\right)^3 = \left(1 - \frac{\alpha z}{r_0}\right)^{-3}.$$

Here $\emptyset_{met}$ is the volume fraction of metal in the mixed metal + silicate plume, $r_0$ is the initial radius of the core of the impactor, *r* is the radius of the descending metal + silicate plume, *z* is the depth of the melt pool or magma ocean and *α* is a constant with the value of ~0.25. For the Grand Tack simulations, Rubie et al. (2015) computed that the fraction of mantle that equilibrates with accreted metal is small: typically, 0.0006–0.014 for planetesimal impacts and 0.008–0.11 for embryo impacts. We have not used the revised hydrodynamic model of Landeau et al. (2021), which includes the effects of impact velocity, because, especially for planetesimal impacts, it predicts much larger volumes of equilibrating mantle than the Deguen et al. model and leads to poor results when fitting Earth's calculated mantle composition to the BSE composition. The application of the Landeau et al. model is discussed in more detail by Dale et al. (2023).

*2.2.2 Dispersed metal-silicate fractionation mechanism.* This mechanism operates when accreted metal becomes suspended and uniformly dispersed as small molten droplets in a vigorously convecting global magma ocean (Fig. 1b). This may happen under three possible scenarios: (i) when an undifferentiated planetesimal impacts an existing magma ocean because the metal is already present as small particles; (ii) when planetesimals are differentiated, but their metallic cores are rapidly emulsified and dispersed as small droplets in a vigorously convecting magma ocean (Kendall and Melosh, 2016); (iii) when the timescale of magma ocean crystallization is short (Elkins-Tanton et al., 2008), so that most planetesimals are accreted onto a solid protoplanet and their metal equilibrates only when the next magma ocean forms as a consequence of a subsequent giant impact (Dale et al., 2023); in this case we expect the metal to become dispersed (Fig. 1b).

The diameter and settling velocity of dispersed metal droplets in a terrestrial magma ocean have been estimated to be ~1 cm and ~0.5 m/s respectively (Rubie et al., 2003). Because convection velocities of deep magma oceans in the hard turbulent regime could be as high as 40 m/s (Solomatov, 2015) we assume that (a) the magma ocean remains chemically well mixed and (b) the metal droplets are kept in suspension and segregate only when they are swept into the mechanical boundary layer at the base of the magma ocean where the vertical component of convection velocities is close to zero (Martin and Nokes, 1988; Höink et al., 2006). Segregation is a continuous process but is modeled by a sequence of discrete iterations until all the dispersed metal has segregated to the core. We define the volume fraction of the boundary layer relative to the whole magma ocean to be ≤0.05. Provided this fraction is less than 0.1, its exact value has only a small effect on results but when it is significantly less than 0.05 the computations become very time-consuming. At each iteration the metal and silicate in the boundary layer are equilibrated using the partitioning/mass balance algorithm of Rubie et al. (2011). The metal is then added to the proto-core assuming that transport occurs in the form of diapirs that sink without further equilibration when the lower part of the mantle is solid (Fig. 1b) and the silicate liquid in the boundary layer is mixed back into the main magma ocean. Iterations are continued until all metal has been extracted from the magma ocean. We assume that the entire segregation process occurs rapidly relative to the decay time scale of $^{182}$Hf to $^{182}$W. The result of this progressive fractionation process is very different to that of a single stage event in which all metal equilibrates with all molten silicate as is often assumed when modelling core formation (see Fig. 7 in Rubie et al., 2003).

In order to apply this model, it is necessary to determine the mass of the global magma ocean as a fraction of the whole mantle. The depth of the base of the magma ocean is determined from the planetesimal equilibration pressure that corresponds to the average fraction of CMB pressures determined by fitting the results to the BSE composition, as explained above.

In Section 3.1 below, we vary the relative fractions of dispersed versus focused planetesimal fractionation events and show that metal delivered by *all* planetesimal impacts needs to equilibrate by the dispersed mechanism in order to satisfy important constraints.

*2.3 Hf-W isotopic evolution.* The Hf-W isotopic system has been incorporated into the accretion/core formation code following the approach adopted by previous studies (Nimmo et al., 2010; Fischer et al., 2018). $^{182}$Hf decays to $^{182}$W with a half-life of 8.90 (±0.09) Myr, i.e. with a decay constant $\boldsymbol{\lambda = 0.077\ \mathrm{Myr^{-1}}}$ (Vockenhuber et al., 2004), whereas $^{180}$Hf and $^{183}$W are stable isotopes. Following previous studies, we define bulk isotopic concentrations at the start of the solar system (time = 0), relative to $^{183}$W: $^{182}$Hf = 2.836 ×10$^{-4}$, $^{182}$W = 1.850664 and $^{183}$W = 1.0 (Jacobson, 2005; Kleine et al., 2009; Nimmo and Agnor, 2006; Nimmo et al., 2010). For $^{180}$Hf/$^{183}$W we use the value of 2.633 which is derived from the results of Kruijer et al. (2014) and Kleine et al. (2004). At each accretional event, the isotopic concentrations in the mantles of the target and impactor bodies are updated using:

$$C^{182Hf} = C_{prev}^{182Hf} exp(-\lambda t)$$

$$C^{182W} = C_{prev}^{182W} + C_{prev}^{182Hf}[1 - exp(-\lambda t)]$$

where $t$ is the time since the previous equilibration event at which concentrations were $C_{\mathrm{prev}}$. After metal-silicate equilibration, when 2 components (e.g. accreted metal and the target's core, or equilibrated and unequilibrated mantle) are homogeneously mixed, the final concentrations $C_{\mathrm{f}}$ of each isotope in the target's mantle and core are given by:

$$C_{\mathrm{f}} = \frac{\sum_{i=1}^{i=2} C_i M_i}{\sum_{i=1}^{i=2} M_i}$$

where $M_i$ are the respective masses and $C_i$ the respective concentrations of the 2 components.

We assume that core-mantle differentiation of embryos and differentiated planetesimals occurred 2 Myr after the start of the solar system based on the timescale of $^{26}$Al heating (Moskovitz and Gaidos 2011; Neumann et al., 2012, Lichtenberg et al., 2018). The start time of N-body simulations should generally correspond to just before the dissipation of the gas disk, and therefore lie in the range 3-8 Myr after the start of the solar system (Haisch et al., 2001; Michel et al., 2021); here we assume a value of 5 Myr. We also show the effects on the W isotope anomaly when the times of differentiation and the start of accretion are changed to (a) 0.5 Myr/5.0 Myr, (b) 2.0 Myr/8.0 Myr, (c) 0.5 Myr/2.0 Myr, and (d) 2.0 Myr/5.0 Myr.

The mantle $^{182}$W anomaly, $\varepsilon_{182W}$, is defined as:

$$\varepsilon_{182W} = \left[\frac{(C^{182W}/C^{183W})}{(C^{182W}/C^{183W})_{CHUR}} - 1\right] \times 10^4$$

where $C^{182W}$ and $C^{183W}$ are mole fraction concentrations of $^{182}$W and $^{183}$W respectively and *CHUR* is the present-day undifferentiated chondritic uniform ratio. For Earth's mantle, $\varepsilon_{182W}$ = 1.9 ±0.1 (Kleine et al., 2002). The mantle Hf/W ratio, expressed as

$$f^{Hf/W} = \left[\frac{(C^{180Hf}/C^{183W})}{(C^{180Hf}/C^{183W})_{CHUR}}\right] - 1$$

is also an important parameter. For Earth's mantle, $f^{Hf/W} = 17 \pm 6$ (Nimmo and Kleine, 2015; Kleine and Walker, 2017), although a higher value of 25.4 ±4.2 has been estimated by Dauphas et al. (2014).

In summary, following each accretion/equilibration event, accreted metal is merged with the target's core and the mantle is homogenized, through convection, by mixing its equilibrated and unequilibrated fractions. The final output of the model is the result of fitting the composition of the BSE for the 13 major and trace oxide and element components $Al_2O_3$, MgO, CaO, FeO, $SiO_2$, Ni, Co, Nb, Ta, V, Cr, W and Mo (König et al., 2011; Palme and O'Neill, 2014; Greber et al., 2015). Five to six free parameters are refined to minimize the reduced chi squared ($\chi^{2)}$) based on a total of 13 elemental constraints. The free parameters are the average pressure of metal-silicate equilibration arising from planetesimal impacts (normalized to the core-mantle-boundary pressure at the time of each fractionation event) and four parameters that define the oxygen content of starting bodies as a function of their heliocentric distance of origin (Fig. 6 in Rubie et al., 2015). A sixth free parameter, either the fraction of the metal of embryo cores that equilibrates (Figs. 2 and S1) or the fraction of dispersed metal-silicate fractionation events (Fig. 3) is also refined in some cases. In addition, calculated values of $f^{Hf/W}$ and $\varepsilon_{182W}$ are compared with BSE values to identify successful models based on a total of 15 constraints.

## 3. Results

*3.1 Variables that control $^{182}W$ isotopic evolution.* As stated above, for modeling Earth's mantle tungsten isotope anomaly ($\varepsilon_{182W}$), in addition to the accretion history that is provided by N-body simulations, we require (i) the fraction of accreted metal that equilibrates with silicate mantle, (ii) the fraction of silicate mantle that equilibrates with accreted metal during each accretion event, and (iii) the tungsten metal-silicate partition coefficient ($D_W$) for each core formation event. In addition, we show below that the timescale of magma ocean crystallization is also a crucial parameter. To constrain these parameters, we first use the Grand Tack N-body simulation 1 (4:1-0.5-8) (Fig. 4 and Fig. 2a in Rubie et al., 2015) which could be representative of Earth's formation (Rubie et al., 2016). This simulates Earth's accretion over a period of 200 Myr and involves seven giant impacts, with the final (Moon-forming) impact occurring 113 Myr after the start of the N-body simulation, followed by the accretion of a small mass (0.3%) of late veneer. 74% of Earth's mass is delivered by giant impacts and 26% by the accretion of planetesimals.

We investigate how the fraction of accreted metal that equilibrates with silicate mantle during embryos' impacts affects the resulting mantle composition and tungsten isotope anomaly. In contrast, we assume that metal delivered by all planetesimals equilibrates completely, because their cores are small (Kendall and Melosh, 2016). Moreover; we assume that in either all or half of the planetesimal impacts the metal-silicate fractionation events occur by the dispersed mechanism. Results (Fig. 2) show that the fit to the BSE improves until the average fraction of embryo cores that equilibrates reaches ~0.6 and remains almost constant beyond this value (Fig. 2a); the latter result is consistent with the findings of Fischer and Nimmo (2018) and Zube et al. (2019). Therefore, for the results presented below we use the value of 0.9 for simulation 1. It is especially noteworthy that the value of $\varepsilon w_{182}$ is almost independent of the fraction of accreted metal from embryo cores that equilibrates (Fig. 2b). This result is contrary to conventional wisdom and is the consequence of only small amounts of silicate liquid equilibrating by the focused mechanism following giant impacts (Fig. 1a). Similar results are obtained for the other five simulations (Fig. S1) and the fractions of equilibrating metal used in each case (usually in the range 0.7-1.0) are listed in Table 1.

Each giant impact causes a temporary drop of $\varepsilon_{182W}$ in the Earth's mantle but this drop is always small and ≤4.9 in $\varepsilon_{180W}$ units for the final giant impacts (Table 1), because of the limited amount of silicate mantle that equilibrates with the impactor's core. This seems to exclude the possibility that Earth accreted within the lifetime of the gas in the protoplanetary disk (which would lead to $\varepsilon_{182W}>15$) and that its final $\varepsilon_{182W}$ was reset by a single late Moon-forming event, as sometimes proposed (Yu and Jacobsen, 2011; Olson & Sharp, 2023; Johansen et al., 2023).

In simulation 6 there is an extremely large giant impact 97 Myr after the start of the simulation with impactor mass/target mass = 0.26. When the average fraction of equilibrating embryo cores is the same for all giant impacts, the best fit chi squared lies in the range 60-70 and the refined pressure at which planetesimal metal equilibrates is close to 0 GPa. When we assume that there is zero fragmentation and metal-silicate equilibration following that specific impact, as predicted by Fig. 8 in Maller et al. (2024) for such a large impactor, we obtain greatly improved results with chi squared = 9.7 (Table 1). In this case, $\varepsilon_{182W}$ of the Earth is affected only by the mixing of the target and impactor's mantles, with their own respective $\varepsilon_{182W}$ values.

The fraction of silicate mantle that equilibrates with accreted metal depends on the relative occurrences of the two metal-silicate fractionation mechanisms described above (Fig. 1a, b). We have investigated the effects of varying the relative occurrences of these two fractionation mechanisms for planetesimal impacts while applying the focused metal-silicate fractionation mechanism (Fig. 1a) for all giant impacts involving impacting embryos. The fraction of dispersed metal-silicate fractionation events for planetesimal impacts is varied from 0 to 1. Using the example of this fraction being 0.4, the computation is performed as follows: for every 10 successive planetesimal impacts in the N-body output, the first 4 are assigned to involve the dispersed mechanism and the following 6 are assigned to involve the focused mechanism.

The dependence on the fraction of planetesimal impacts causing dispersed metal-silicate fractionation events of the calculated final concentration of W in the mantle, the quality of the global fit to BSE chemistry (expressed as $\chi^2$), $f^{Hf/W}$ and the tungsten isotopic anomaly $\varepsilon_{182W}$ are shown in Fig. 3. These results show that this fraction must be close to 1.0 in order to satisfy the BSE W concentration constraint and to minimize $\chi^2$. Although the best fit W concentration is still slightly high, it matches the BSE value well within the uncertainties (Fig. 3a). A recent study suggests that the BSE W concentration could be as high as 27-31 ppb (Peters et al., 2023). We do not use this estimate because it is based on lithospheric mantle xenoliths, of which the W inventory is metasomatically overprinted. The strong dependence in $\chi^2$ (Fig. 3b) is caused almost entirely by the effects of fractionation mechanisms on final mantle concentrations of W and Mo (Fig. S2 shows a typical Mo trend). Compared with assuming that all starting bodies were equally enriched in refractory elements (Rubie et al., 2015), the new restriction of enhanced refractory element starting concentrations to highly-reduced bodies originating at less than around 1 au also helps to decrease the final calculated W and Mo concentrations. This is due to more of these elements being fractionated into metal due to the highly-reducing conditions. Omitting W and Mo concentrations from the determination of $\chi^2$ results in values that are consistently low (i.e. good) irrespective of the relative occurrences of the two metal-silicate fraction mechanisms (Dale et al. 2023). This emphasizes the importance of fitting as many siderophile elements as possible in core formation studies. Finally, both calculated $f^{\text{Hf/W}}$ and $\varepsilon_{182W}$ values are also affected strongly by the relative occurrences of the two planetesimal metal-silicate fractionation mechanisms and only become consistent with Earth's mantle values when the dispersed fractionation mechanism dominates strongly. Fig. 3d shows that the relative occurrences of metal-silicate fractionation mechanisms are just as important as the chronology of accretion and core formation in determining the final $\varepsilon_{182W}$ value. However, this does not mean that the chronology of Earth's accretion is unconstrained by the $^{182}$W anomaly, because the relative occurrence of the dispersed metal-silicate fractionation mechanism is constrained independently by the W concentration and $f^{\text{Hf/W}}$ (Fig. 3a and 3c). Also shown in Fig. 3d are the effects of changing the assumed times at which starting bodies differentiated and accretion started.

The result that most or all planetesimal metal-silicate fraction events occurred by the dispersed mechanism does not, in itself, distinguish between the possibilities that either (i) the timescale of magma ocean crystallization is long, but most impacting bodies are undifferentiated or their cores emulsify rapidly or (ii) the timescale of magma ocean solidification is short, so that planetesimals (differentiated or not) impact a solid proto-planet and equilibrate only when a new magma ocean is created by the next giant impact. However, because the times of magma ocean crystallization are often short compared with the time intervals between giant impacts (Elkin-Tanton, 2008; Supplementary Material), the "instantaneous magma ocean solidification" model is likely to be closest to reality. Results for the "continuous magma ocean" model, according to which metal delivered by each planetesimal equilibrates at the time of its impact, are shown in Fig. 4a. Results for the "instantaneous magma ocean solidification" model, according to which all planetesimals are accreted to a solid Earth with the metal only equilibrating later in the magma ocean that is created by the subsequent giant impact, are shown in Fig. 4b. In both cases, the black symbols/lines show the original N-body accretion history and the red symbols/lines show the N-body timescale adjusted in order the achieve $\varepsilon_{182W}$= 1.9 for the two respective models (as justified below). The results are dramatically different for the two magma ocean models: the time of the final giant impact is predicted to be 92 Myr for the continuous magma ocean model and 54 Myr for the instantaneous magma ocean solidification model. This is because in the instantaneous magma ocean solidification case the W delivered by planetesimals resides in the Earth's mantle for longer (i.e. until the next global magma ocean forms) than in the case where planetesimals' metal equilibrates immediately upon impact, as in the continuous magma ocean case.

The metal-silicate partition coefficient of W, $D_W^{m/s} = X_W^{metal}/X_{WO_{n/2}}^{silicate}$ where $X$ is mole fraction and $n$ is the valence of W in silicate liquid, is an important parameter which in some previous Hf-W studies has been assumed to have a constant value throughout accretion and core formation (Rudge et al., 2010; Zube et al., 2019). In addition to $P$ and $T$, this parameter is strongly affected by the carbon content of the metal and oxygen fugacity (O'Neill et al., 2008; Cottrell et al., 2009; Wade et al., 2012; Jennings et al., 2021). For each core-forming event we determined $D_W^{m/s}$ using the parameterization of Jennings et al. (2021). Results show that $D_W^{m/s}$ varies by several orders of magnitude during accretion, from ~12000 to ~70 (Fig. Fig. 5a). The highest values occur very early during accretion when oxygen fugacities of metal-silicate equilibration are very low (~4 log units below the iron-wüstite (IW) buffer), whereas during the later stages conditions are more oxidizing (~2 log units below IW) and result in much lower $D_W^{m/s}$ values (Fig. 5b, see also Fischer and Nimmo, 2018). Such an evolution of oxygen fugacity is well established by core formation studies with the early accretion of highly-reduced bodies originating in the innermost solar system at <1-1.5 au and the later accretion of more oxidized bodies originating from greater heliocentric distances (e.g., Rubie et al., 2011, 2015; Wade and Wood, 2005; Wood et al., 2006).

*3.2 Evaluation of six Grand Tack simulations of Earth's accretion.*

We re-evaluate the results of the six Grand Tack simulations considered by Rubie et al. (2015) using our revised accretion/core formation model. As described above, modifications include quantification of the depth of giant-impact-induced melting from SPH simulations, inclusion of the new dispersed metal-silicate fractionation mechanism for planetesimal impacts, the restriction of enhanced refractory element concentrations to highly-reduced starting bodies that originate at less than ~1 au, and the inclusion of the Hf-W isotopic system. Following Dale et al. (2023) and the results of Fig. 3 we assume in the following that the dispersed metal-silicate fractionation mechanism operates for *all* planetesimal impacts.

Our new results obtained for the 6 simulations of Rubie et al. (2015) are presented in Table 1 and Fig. 6. In Table 1, we list results for both the continuous magma ocean model and the instantaneous magma ocean solidification model, although neither is likely to be fully realistic. Elkins-Tanton (2008) concluded that magma ocean solidification times range from <100,000 yr for volatile-poor magma oceans and < 5 Myr for volatile-rich magma oceans. Here, we consider the case of all magma oceans solidifying after 5 Myr which we term the "magma ocean lifetime of 5 Myr model". For the first 5 Myr after each giant impact and prior to the subsequent giant impact, the metal accreted by planetesimals equilibrates with silicate liquid at the time of each impact. At times >5 Myr after a giant impact, accreted metal equilibrates later (sometimes 10's of Myr later) in the magma ocean that is created by the subsequent giant impact. Based on the accretion timescale of simulation 3 (supplementary material) this is a very simplistic because only a single time interval between giant impacts exceeds 5 Myr. However, because this time interval is very long (~160 Myr) the model results are still different from those of the continuous magma ocean model. Reality must lie between our "instantaneous solidification" and "lifetime of 5 Myr" models. It is likely that magma oceans that formed early, when the proto planet was small and volatile-poor, would have solidified in times <10,000 yr (Elkins-Tanton, 2008).

We judge the success of each simulation by comparing calculated mantle values of W and Mo concentrations, $f^{Hf/W}$, and the $^{182}W$ isotope anomaly with BSE values. In Table 1, calculated values that are inconsistent with BSE values are shown in red text.

In some simulations, the calculated value of the $^{182}W$ isotope anomaly is too high and in others it is too low (Table 1, Fig. 6), indicating that accretion timescales are either too fast or too slow. Thus, as in Fig. 4, we rescale the timescale of each simulation by multiplying all times by a factor *t-adjust*, and determine which value of this parameter reproduces the BSE value of $\varepsilon_{182W}$ of 1.9±0.1. While this is, of course, completely artificial, it helps to gain insight into the actual Earth accretion timescale, even if the simulations did not reproduce it self-consistently. This timescale correction affects only $\varepsilon_{182W}$ and all other compositional results are unaffected.

Only the results of simulation 3 are consistent with all four BSE criteria (Table 1, supplementary material). For this simulation, the instantaneous crystallization magma ocean model results in $\varepsilon_{182W}$

= 2.0 and, with magma ocean lifetimes of 5 Myr, $\varepsilon_{182W}$ = 2.6. In contrast, the continuous magma ocean model predicts $\varepsilon_{182W}$ = 3.7. This simulation is therefore consistent with average magma ocean lifetimes <5 Myr. According to the instantaneous crystallization model, the final giant impact occurs 143 Myr after the start of the solar system which is identical to an estimated late age of Moon formation of 142 ± 25 Myr (Maurice et al., 2020); however, this age is currently controversial (Kruijer et al., 2021, Thiemens et al., 2019, 2021). With magma ocean lifetimes of 5 Myr, the final giant impact occurs even later, 183 Myr after the start of the solar system.

Simulation 4 fails to reproduce the four criteria and we do not consider it further in Table 1 or Fig. 6. If we relax the constraint imposed by the BSE Mo concentration, simulations 1, 2, 4 and 6 can be considered if the accretion timescales are artificially adjusted to give $\varepsilon_{182W}$ = 1.9 (Table 1, Fig. 6). These adjustments, assuming end-member magma ocean lifetimes of 0 and 5 Myr, result in final giant impacts occurring 53-62 Myr in simulations 1, 2 and 5 and 134-168 Myr in simulation 6 after the start of the solar system (Table 1, Fig. 6).

Although the young age for the Moon estimated for simulation 3 is consistent with the results of Maurice et al. (2020), the results of isotopic (Barboni et al., 2017; Thiemens et al., 2019; Greer et al., 2023) and modelling (Woo et al., 2024) studies are similar to the earlier ages that are predicted by simulations 1, 2 and 5. Therefore, our results demonstrate that an accurate determination of the age of the Moon would provide an additional constraint on the validity of accretion simulations.

## 4. Conclusions

Major developments to the accretion/core-mantle differentiation model of Rubie et al. (2015) presented here include (i) the use of SPH simulation results to estimate the depth of melting caused by giant impacts, (ii) the application of a new metal-silicate fractionation model (Fig. 1b) for planetesimal accretion events, and (iii) incorporation of the Hf-W isotope system to constrain the timescales of accretion. Results show that the timescale of magma ocean crystallization has a major effect on the mantle $^{182}$W anomaly. Only one of the six simulations studied by Rubie et al. (2015) fully satisfies all the constraints imposed by BSE concentrations of W and Mo, $f^{Hf/W}$ and $\varepsilon_{182W}$ and is consistent with magma ocean crystallization timescales being ≤5 Myr and that the Moon-forming giant impact occurred late, 143-183 Myr after the start of the solar system. Other simulations, however, predict much earlier Moon-forming impacts occurring 53-62 Myr after the start of the solar system.

This revised model has great potential for identifying viable models of Earth's accretion, especially when mantle chemistry alone is not discriminative (e.g., Rubie et al., 2015; Dale et al., 2023; Gu et al., 2023). Future improvements could include using the age of the Moon as an additional constraint on the validity of accretion simulations when current uncertainties have been resolved (Kruijer et al. 2021; Thiemens et al. 2019, 2021) and incorporating modeled estimates of magma ocean solidification timescales.

**CRediT authorship contribution statement**

**David C. Rubie:** Conceptualization, Investigation, Methodology, Software, Validation, Visualization, Writing – original draft, Writing – review & editing. **Katherine I. Dale:** Methodology, Software, Writing – review & editing. **Gabriel Nathan:** Methodology, Writing – review & editing, Software. **Miki Nakajima:** Methodology, Writing – review & editing. **Eleanor S. Jennings:** Methodology, Writing – review & editing. **Gregor J. Golabek:** Methodology, Software, Writing – review & editing. **Seth A. Jacobson:** Software, Writing – review & editing. **Alessandro Morbidelli:** Conceptualization, Methodology, Software, Writing – review & editing.

**Data availability**

Additional data and software are available on request.

**Acknowledgments**

This work has been supported by the ERC project HolyEarth (No. 101019380). M.N. was supported in part by the National Aeronautics and Space Administration (NASA) grant numbers 80NSSC19K0514 and 80NSSC21K1184. Partial funding for M.N. was also provided by NSF EAR-2237730 as well as the Center for Matter at Atomic Pressures (CMAP), an NSF Physics Frontier Center, under Award PHY-2020249. M.N. was also supported in part by the Alfred P. Sloan Foundation under grant G202114194. We thank Thorsten Kleine and Francis Nimmo for discussions and Carsten Münker and an anonymous reviewer for their helpful and detailed reviews.

**References**

Barboni, M., Boehnke, P., Keller, B., et al., 2017. Early formation of the Moon 4.51 billion years ago. Sci. Adv. 3, e1602365.

Blanchard, I., Rubie, D.C., Jennings E.S., Franchi, I.A., Zhao, X., Petitgirard, S., Miyajima, N., Jacobson, S.A., Morbidelli, A., 2022. The metal–silicate partitioning of carbon during Earth's accretion and its distribution in the early solar system. Earth Planet. Sci. Lett. 580, 117374.

Cottrell, E., Walter, M.J., Walker, D., 2009. Metal–silicate partitioning of tungsten at high pressure and temperature: implications for equilibrium core formation in Earth. Earth Planet. Sci. Lett. 281, 275–287.

Dale, K.I., Rubie, D.C., Nakajima, M., Jacobson, S., Nathan, G., Golabek, G.J., Cambioni, S., Morbidelli, A., 2023. An improved model of metal/silicate differentiation during Earth's accretion. Icarus, https://doi.org/10.1016/j.icarus.2023.115739.

Dauphas, N., Burkhardt, C., Warren, P.H., Teng, F.-Z., 2014. Geochemical arguments for an Earth-like Moon-forming impactor. Philos. Trans. R. Soc. A372, 20130244.

Deguen, R., Olson, P., Cardin, P., 2011. Experiments on turbulent metal-silicate mixing in a magma ocean. Earth Planet. Sci. Lett. 310, 303-313.

Deguen, R., Landeau, M., Olson, P., 2014. Turbulent metal–silicate mixing, fragmentation, and equilibration in magma oceans. Earth Planet. Sci. Lett. 391, 274-287.

Elkins-Tanton, L.T., 2008. Linked magma ocean solidification and atmospheric growth for Earth and Mars. Earth Planet. Sci. Lett. 271, 181-191.

Fischer, R.A., Campbell, A.J., Ciesla, F.J., 2017. Sensitivities of Earth's core and mantle compositions to accretion and differentiation processes. Earth Planet. Sci. Lett. 458, 252-262.

Fischer, R.A., Nimmo, F., 2018. Effects of core formation on the Hf–W isotopic composition of the Earth and dating of the Moon-forming impact. Earth Planet. Sci. Lett. 499, 257-265.

Greber, N.D., Puchtel, I.S., Nägler, T.F., Mezger, K., 2015. Komatiites constrain molybdenum isotope composition of the Earth's mantle. Earth Planet. Sci. Lett. 421, 129–138.

Greer, J., Zhang, B., Isheim, D., Seidman, D.N., Bouvier, A., Heck, P.R., 2023. 4.46 Ga zircons anchor chronology of lunar magma ocean. Geochem. Persp. Let. 27, 49-53.

Gu, J.T., Fischer, R.A., Brennan, M.C., Clement, M.S., Jacobson, S.A., Kaib, N.A., O'Brien, D.P., Raymond, S.N., 2023. Comparisons of the core and mantle compositions of earth analogs from different terrestrial planet formation scenarios. Icarus 394, 115425.

Haisch, J., Karl, E., Lada, E. A., Lada, C. J., 2001. Disk frequencies and lifetimes in young clusters. Astrophys. J. Lett. 553, L153–L156.

Hansen, B.M.S. 2009. Formation of the Terrestrial Planets from a Narrow Annulus. The Astrophysical Journal 703, 1131–1140. doi:10.1088/0004-637X/703/1/1131.

Hirschmann, M.M., 2018. Comparative deep Earth volatile cycles: The case for C recycling from exosphere/mantle fractionation of major ($H_2O$, C, N) volatiles and from $H_2O$/Ce, $CO_2$/Ba, and $CO_2$/Nb exosphere ratios. Earth Planet. Sci. Lett. 502, 262–273, https://doi.org/10.1016/j.epsl.2018.08.023.

Hirschmann, M.M., Bergin, E.A., Blake, G.A., Ciesla, F.J., Li, J., 2021. Early volatile depletion on planetesimals inferred from C–S systematics of iron meteorite parent bodies. Proc. Natl. Acad. Sci. U.S.A. 118, 1–8.

Höink, T., Schmalzl, J., Hansen, U., 2006. Dynamics of metal-silicate separation in a terrestrial magma ocean. Geochem. Geophys. Geosyst. 7, Q09008, doi:10.1029/2006GC001268.

Jacobsen, S.B., 2005. The Hf-W isotopic system and the origin of the Earth and Moon. *Annu. Rev. Earth Planet. Sci*. 33, 531–70.

Jennings, E.S., Jacobson, S.A., Rubie, D.C., Nakajima, Y., Vogel, A.K., Rose-Weston, L.A., Frost, D.J., 2021. Metal-silicate partitioning of W and Mo and the role of carbon in controlling their abundances in the bulk silicate earth. Geochim. Cosmochim. Acta 293, 40-69, doi.org/10.1016/j.gca.2020.09.035.

Johansen, A., Ronnet, T., Schiller, M., Deng, Z., Bizzarro, M., 2023. Anatomy of rocky planets formed by rapid pebble accretion. II. Differentiation by accretion energy and thermal blanketing. Astronomy and Astrophysics 671. doi:10.1051/0004-6361/202142142.

Kendall, J.D., Melosh, H.J., 2016. Differentiated planetesimal impacts into a terrestrial magma ocean: Fate of the iron core. Earth Planet. Sci. Lett. 448, 24-33.

Kleine, T., Münker, C., Mezger, K., Palme, H., 2002. Rapid accretion and early core formation on asteroids and the terrestrial planets from Hf–W chronometry. Nature 418, 952–955.

Kleine, T., K. Mezger, K., Palme, H., Münker, C., 2004. The W isotope evolution of the bulk silicate Earth: constraints on the timing and mechanisms of core formation and accretion Earth Planet. Sci. Lett. 228, 109-123.

Kleine, T., Touboul, M.,, Bourdon, B., Nimmo, F., Mezger, K., Palm, H., Jacobsen, S.B., Yin, Q.-Z., Halliday, A.N., 2009. Hf–W chronology of the accretion and early evolution of asteroids and terrestrial planets. Geochim. Cosmochim. Acta 73, 5150–5188.

Kleine, T., Walker, R.J., 2017. Tungsten isotopes in planets. Annu. Rev. Earth Planet. Sci. 45, 389–417

König, S., Münker, C., Hohl, S., Paulick, H., Barth, A.R., Lagos, M., Pfänder, J., Büchl, A., 2011. The Earth's tungsten budget during mantle melting and crust formation. Geochim. Cosmochim. Acta 75, 2119-2136.

Maller, A., Landeau, M., Allibert, L., Charnoz, S., 2024. Condition for metal fragmentation during Earth-forming collisions. Phys. Earth Planet. Int. 352, 107199.

Maurice, M., Tosi, N., Schwinger, S., Breuer, D., Kleine, T., 2020. A long-lived magma ocean on a young Moon. Sci. Adv. 6, eaba8949.

Nesvorný, D., Roig, F.V., Deienno, R. 2021. The Role of Early Giant-planet Instability in Terrestrial Planet Formation. The Astronomical Journal 161. doi:10.3847/1538-3881/abc8ef.

Nimmo, F., Kleine, T., 2015. Early Differentiation and Core Formation: Processes and Timescales. In: *The Early Earth: Accretion and Differentiation* (eds. J. Badro, M. Walter). Geophysical Monograph Series, American Geophysical Union, pp. 82-202, DOI:10.1002/9781118860359.

Kruijer, T.S., Kleine, T., Fischer-Gödde, M.F., Burkhardt, C., Wieler, R., 2014. Nucleosynthetic W isotope anomalies and the Hf–W chronometry of Ca–Al-rich inclusions. Earth Planet. Sci. Lett. 404, 317-327.

Kruijer, T.S., Archer, G.J., Kleine, T., 2021. No $^{182}$W evidence for early Moon formation. Nat. Geosci. 14, 714–715.

Landeau, M., Deguen, R., Phillips, D., Neufeld, J.A., Lherm, V., Dalziel, S.B., 2021. Metal-silicate mixing by large Earth-forming impacts. Earth Planet. Sci. Lett. 564, 116888.

Lichtenberg, T., Golabek, G.J., Dullemond, C.P., Schönbächler, M., Gerya, T.V., Meyer, M.R., 2018. Impact splash chondrule formation during planetesimal recycling. Icarus 302, 27-43.

Martin, D., Nokes, R., 1988. Crystal settling in a vigorously convecting magma chamber. Nature 332, 534-536.

Michel, A., van der Marel, N. , Matthews, B.C., 2021. Bridging the Gap between Protoplanetary and Debris Disks: Separate Evolution of Millimeter and Micrometer-sized Dust. The Astrophysical Journal 92, 72.

Monteux, J., Golabek, G.J., Rubie, D.C., Tobie, G., Young, E.D., 2018. Water and the interior structure of terrestrial planets and icy bodies. Space Science Reviews 214:39.

Morbidelli, A., Libourel, G., Palme, H., Jacobson, S.A., Rubie, D.C. 2020. Subsolar Al/Si and Mg/Si ratios of non-carbonaceous chondrites reveal planetesimal formation during early condensation in the protoplanetary disk. Earth Planet. Sci. Lett. 538.

Moskovitz, N., Gaidos, E., 2011. Differentiation of planetesimals and the thermal consequences of melt migration. Meteoritics and Planetary Science 46, 903-918.

Nakajima, M., Golabek, G.J., Wünnemann, K., Rubie, D.C., Burger, C., Melosh, H.J., Jacobson, S.A., Manske, L., Hull, S.D., 2021. Scaling laws for the geometry of an impact-induced magma ocean. Earth Planet. Sci. Lett. 568, 116983.

Neumann, W., Breuer, D., Spohn, T., 2012. Differentiation and core formation in accreting planetesimals. A&A 543, A141.

Nimmo, F., Agnor, C.B., 2006. Isotopic outcomes of N-body accretion simulations: Constraints on equilibration processes during large impacts from Hf/W observations. Earth Planet. Sci. Lett. 243, 26-43.

Nimmo, F., O'Brien, D.P., Kleine, T., 2010. Tungsten isotopic evolution during late-stage accretion: Constraints on Earth–Moon equilibration. Earth Planet. Sci. Lett. 292, 363-370.

O'Neill, H. S. C., Berry A. J., Eggins S. M., 2008. The solubility and oxidation state of tungsten in silicate melts: Implications for the comparative chemistry of W and Mo in planetary differentiation processes. Chem. Geol. 255, 346-359.

Olson, P.L., Sharp, Z.D., 2023. Hafnium-tungsten evolution with pebble accretion during Earth formation. Earth Planet. Sci. Lett. 622, 118418.

Palme, H., O'Neill, H.S.C., 2014. Cosmochemical Estimates of Mantle Composition. In: *Treatise on Geochemistry (Second Edition)* (eds. H. D. Holland and K. K. Turekian). Elsevier, Oxford, pp. 1-39.

Peters, D., Rizo, H., Carlson, R.W., Walker, R.J., Rudnick, R.L., Luguet, A., 2023. Tungsten in the mantle constrained by continental lithospheric peridotites: Less incompatible and more abundant. Geochim. Cosmochim. Acta 351, 167-180.

Raymond, S.N., Kokubo, E., Morbidelli, A., Morishima, R., Walsh, K., 2014. Terrestrial Planet Formation at Home and Abroad. In *Protostars and Planets* VI, 595–618. 10.2458/azu_uapress_9780816531240-ch026.

Rubie, D.C., Melosh, H.J., Reid, J.E., Liebske, C., Righter, K., 2003. Mechanisms of metal-silicate equilibration in the terrestrial magma ocean. Earth Planet. Sci. Lett. 205, 239-255.

Rubie, D.C., Frost, D.J., Mann, U., Asahara, Y., Tsuno, K., Nimmo, F., Kegler, P., Holzheid, A., Palme, H., 2011. Heterogeneous accretion, composition and core-mantle differentiation of the Earth. Earth Planet. Sci. Lett. 301, 31-42, doi: 10.1016/j.epsl.2010.11.030.

Rubie, D.C., Jacobson, S.B., Morbidelli, A., O'Brien, D.P., Young, E.D., de Vries, J., Nimmo, F., Palme, H., Frost, D.J., 2015. Accretion and differentiation of the terrestrial planets with implications for the compositions of early-formed Solar System bodies and accretion of water. Icarus 248, 89-108.

Rubie, D.C., Laurenz, V., Jacobson, S.A., Morbidelli, A., Palme, H., Vogel, A.K., Frost, D.J., 2016. Highly siderophile elements were stripped from Earth's mantle by iron sulfide segregation. Science 353, 1141-1144.

Rudge, J.F., Kleine, T., Bourdon, B., 2010. Broad bounds on Earth's accretion and core formation constrained by geochemical models. Nature Geoscience 3, 439-443.

Solomatov, V. S., 2015. Magma oceans and primordial mantle differentiation. In *Treatise on Geophysics, 2nd Edition*, edited by G. Schubert, v. 9, Elsevier, pp. 81-104.

Thiemens, M.M., Sprung, P., Fonseca, R.O.C. et al., 2019. Early Moon formation inferred from hafnium–tungsten systematics. Nat. Geosci. 12, 696–700.

Thiemens, M.M., Tusch, J., Fonseca, R.O.C. et al., 2021. Reply to: No $^{182}$W evidence for early Moon formation. Nature Geosci. 14, 716-718.

Vockenhuber, C., Oberli, F., Bichler, M., Ahmad, I., Quitté, G., Meier, M., Halliday, A.N. Lee, D.-C., Kutschera, W., Steier, P., Gehrke, R.J., Helmer, R.G., 2004. New Half-Life Measurement of $^{182}$Hf: Improved Chronometer for the Early Solar System. Phys. Rev. Lett. 93, 172501.

Wade, J., Wood, B.J., 2005. Core formation and the oxidation state of the Earth. Earth Planet. Sci. Lett. 236, 78-95.

Wade. J., Wood, B.J., Tuff. J., 2012. Metal–silicate partitioning of Mo and W at high pressures and temperatures: Evidence for late accretion of sulphur to the Earth. Geochim. Cosmochim. Acta 85, 58–74.

Woo, J.M.Y., Nesvorný, D., Scora, J., Morbidelli, A., 2024. Terrestrial planet formation from a ring: Long-term simulations accounting for the giant planet instability. Icarus 417. doi:10.1016/j.icarus.2024.116109.

Walsh, K., Morbidelli, A., Raymond, S. et al., 2011. A low mass for Mars from Jupiter's early gas-driven migration. Nature 475, 206–209.

Yu, G., Jacobsen, S.B., 2011. Fast accretion of the Earth with a late Moon-forming giant impact. Proc. Natl Acad. Sci.108, 17604–17609.

Wood, B.J., Walter, M.J., Wade, J., 2006. Accretion of the Earth and segregation of its core. Nature 441, 825-833, doi:10.1038/nature04763.

Zube, N.G., Nimmo, F., Fischer, R.A., Jacobson, S.A., 2019. Constraints on terrestrial planet formation timescales and equilibration processes in the Grand Tack scenario from Hf-W isotopic evolution. Earth Planet. Sci. Lett. 522, 210-218.

# Table 1
Summary of best fit results for Earth-like planets.

| Simulation # | sim_1<br>4:1-0.5-8* | sim_2<br>4:1-0.25-7 | sim_3<br>i4:1-0.8-6 | sim_4<br>8:1-0.8-8 | sim_5<br>8:1-0.25-2 | sim_6<br>i4:1-0.8-4 | Earth's mantle |
|---|---|---|---|---|---|---|---|
| Initial mass of each embryo | $5.0 \times 10^{-2} M_e$ | $2.5 \times 10^{-2} M_e$ | $2.5 \times 10^{-2}$ to $7.7 \times 10^{-2} M_e$ | $8.0 \times 10^{-2} M_e$ | $2.5 \times 10^{-2} M_e$ | $2.5 \times 10^{-2}$ to $7.7 \times 10^{-2} M_e$ | |
| Mass of each planetesimal | $3.8 \times 10^{-4} M_e$ | $3.8 \times 10^{-4} M_e$ | $3.8 \times 10^{-4} M_e$ | $3.8 \times 10^{-4} M_e$ | $3.8 \times 10^{-4} M_e$ | $3.8 \times 10^{-4} M_e$ | |
| **Results for model Earths:** | | | | | | | |
| Mass fraction of accreted embryos | 0.74 | 0.67 | 0.75 | 0.90 | 0.84 | 0.69 | |
| Mass fraction of accreted planetesimals | 0.26 | 0.33 | 0.25 | 0.10 | 0.16 | 0.31 | |
| Time of final giant impact (Myr) [1] | 118 | 103 Myr | 136 Myr | 85 Myr | 33 Myr | 112 Myr | |
| Chi squared [2] | 4.9 | 4.2 | 1.8 | **11.0** | 7.4 | 9.7 | |
| Average fraction of embryo cores that eqilibrates | 0.9 | 1.0 | 0.8 | 0.7 | 0.8 | 0.9** | |
| Mantle W (ppb) | 15.8±2.6 | 15.6±4.2 | 11.7±5.7 | **23.0±4.8** | 15.4±5.8 | 9.4±4.2 | 12.0±3.6 |
| Mantle Mo (ppb) | **67.1±18.3** | **48.0±13.2** | 46.3±20.2 | **74.5±20.0** | **71.4±18.9** | **47.3±9.5** | 23±7 |
| $f^{Hf/W}$ | 11.8 | 11.8 | 16.4 | **7.8** | 12.4 | 21.0 | 17±6 |
| ***Continuous magma ocean model:*** | | | | | | | |
| $\varepsilon_{182W}$ [3] | **1.3** | **0.9** | **3.7** | **2.6** | **4.9** | **5.0** | 1.9±0.1 |
| t_adjust | 0.78 | 0.67 | 1.60 | | 2.05 | 2.10 | |
| Recalculated time of last giant impact | 92 Myr | 69 Myr | 217 Myr | | 68 Myr | 235 Myr | |
| Change in $\varepsilon_{182W}$ caused by last giant impact [4] | -4.9 | -0.85 | -0.05 | | +0.05 | +0.33 | |
| ***Instantaneous crystallization magma ocean model:*** | | | | | | | |
| $\varepsilon_{182W}$ [3] | **0.4** | **0.5** | 2.0 | | **3.8** | **2.4** | 1.9±0.1 |
| t_adjust | 0.46 | 0.51 | 1.05 | | 1.65 | 1.2 | |
| Recalculated time of last giant impact | 54 Myr | 53 Myr | 143 Myr | | 55 Myr | 134 Myr | |
| Change in $\varepsilon_{182W}$ caused by last giant impact [4] | -4.3 | -0.7 | -0.06 | | -0.02 | -0.14 | |
| ***Magma ocean lifetime of 5 Myr model:*** | | | | | | | |
| $\varepsilon_{182W}$ [3] | **0.5** | **0.6** | **2.6** | | **4.3** | **3.1** | 1.9±0.1 |
| t_adjust | 0.52 | 0.54 | 1.35 | | 1.88 | 1.50 | |
| Recalculated time of final giant impact | 61 Myr | 56 Myr | 183 Myr | | 62 Myr | 168 Myr | |
| Change in $\varepsilon_{182W}$ caused by last giant impact [4] | -4.7 | -0.43 | -0.05 | | +0.01 | -0.14 | |

*Simulation names convey information about the starting disk parameters. For example, in 4:1-0.5-8, "4:1" indicates the total starting mass of embryos ratioed to that of planetesimals, the second parameter

indicates the initial mass of the largest embryo (0.025, 0.05 or 0.08 $M_e$) and "8" is the run number. The prefix "i" indicates that the initial mass of embryos increases with heliocentric distance.

<span style="color:red">Values are listed in red when they are inconsistent with BSE values</span>

[1] Time of final giant impact according to the N-body simulation when starting at 5 Myr

[2] Chi squared values are based on fitting calculated mantle chemistry to BSE concentrations of MgO, FeO, SiO$_2$, Ni, Co, Nb, Ta, V, Cr, W (König et al., 2011 Palme and O'Neill, 2014), and Mo (Greber et al., 2015).

[3] $\varepsilon_{182W}$ is based on the N-body timescale and t_adjust is the factor by which the N-body timescale is multiplied in order to obtain $\varepsilon_{182W}$ = 1.9.

[4] Based on the recalculated timescales required to achieve $\varepsilon_{182W}$ = 1.9.

** In simulation 6 there is an extremely large giant impact at 97 Myr with impactor mass/target mass = 0.26. Because of this large ratio, we assume that there is no fragmentation or metal-silicate equilibration (Fig. 8 in Maller et al. 2024).

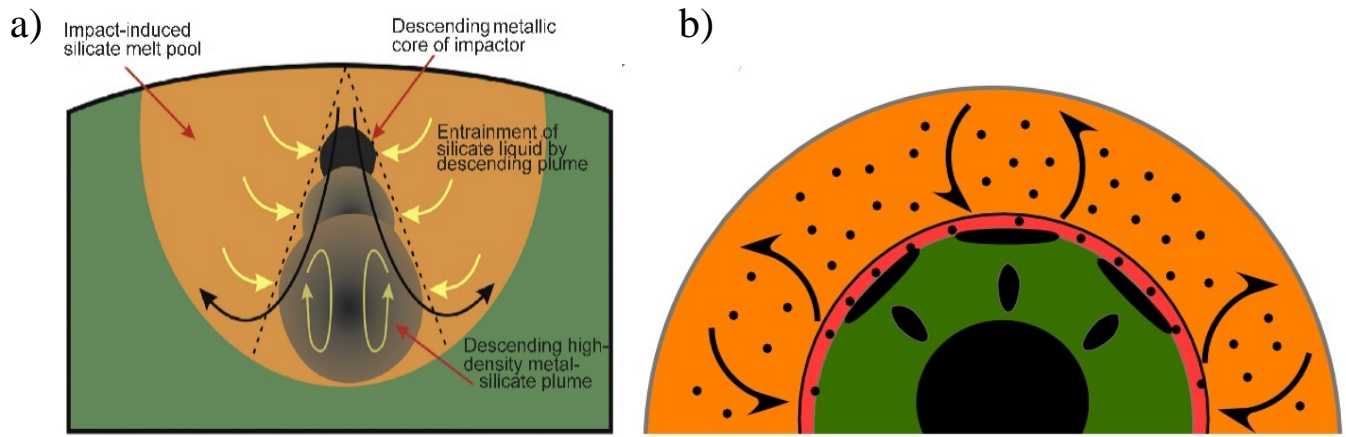

**Fig. 1.** Metal-silicate fractionation models. a) *Focused metal-silicate fractionation*. When a large impacting body is differentiated, as is the case here, its metallic core sinks in the impact-induced silicate-magma melt pool (or global magma ocean), and progressively emulsifies and entrains increasing amounts of the surrounding silicate liquid. The result is a high-density plume of molten metal + molten silicate that expands as it sinks to the base of the melt pool. Final chemical equilibration of metal and silicate liquids occurs within this plume when it reaches the bottom of the melt pool/magma ocean. (After Rubie et al., 2015.) b) *Dispersed metal-silicate fractionation*. Cross section of an accreting planet showing a global magma ocean, overlying crystalline mantle and the proto-core. Small (e.g. ~1-10 mm) metal droplets (black) are suspended in the turbulently convecting magma ocean and only segregate when they are swept into the mechanical boundary layer (red) at the base of the magma ocean (Martin and Nokes, 1988; Höink et al., 2006). The segregating metal finally equilibrates with the silicate liquid in this boundary layer before accumulating and descending to the core as sinking diapirs (black) without further equilibration. The silicate liquid in the boundary layer is continuously mixed with the bulk of the magma ocean during the segregation process, as the result of rapid turbulent convection. Crystalline mantle: green; magma pond/ocean: orange; core and core-forming metal: black.

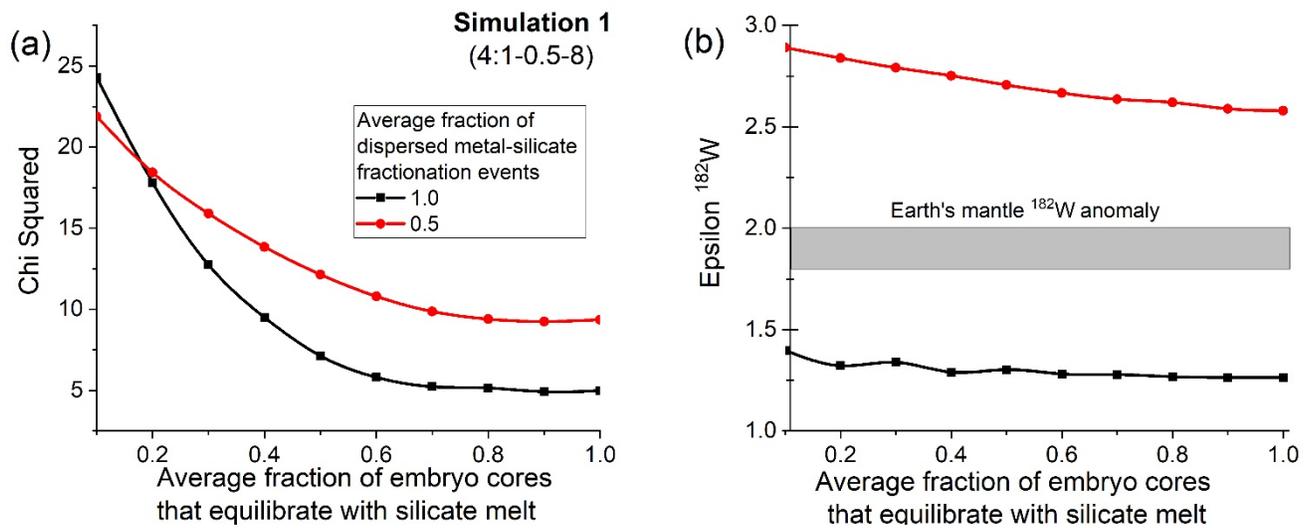

**Fig. 2.** The effects on (a) chi squared and (b) $\varepsilon_{182W}$ of varying the average fraction of embryo cores that equilibrates with silicate liquid at the base of giant impact-induced melt pools in simulation 1. Reduced chi squared is determined from fitting calculated mantle concentrations of MgO, FeO, $SiO_2$, Ni, Co, Nb, Ta, V, Cr, W and Mo to BSE values (Palme and O'Neill, 2014; Greber et al., 2015). Results are shown for fractions of dispersed metal-silicate fractionation events for planetesimal impacts equal to 1.0 (black curves) and 0.5 (red curves). The fraction of dispersed metal-silicate fractionation events has a large effect on chi squared and $\varepsilon_{182W}$ mainly through its effect on calculated mantle concentrations of W (Fig. 3). Results for simulations 2-6 are shown in Fig. S1.

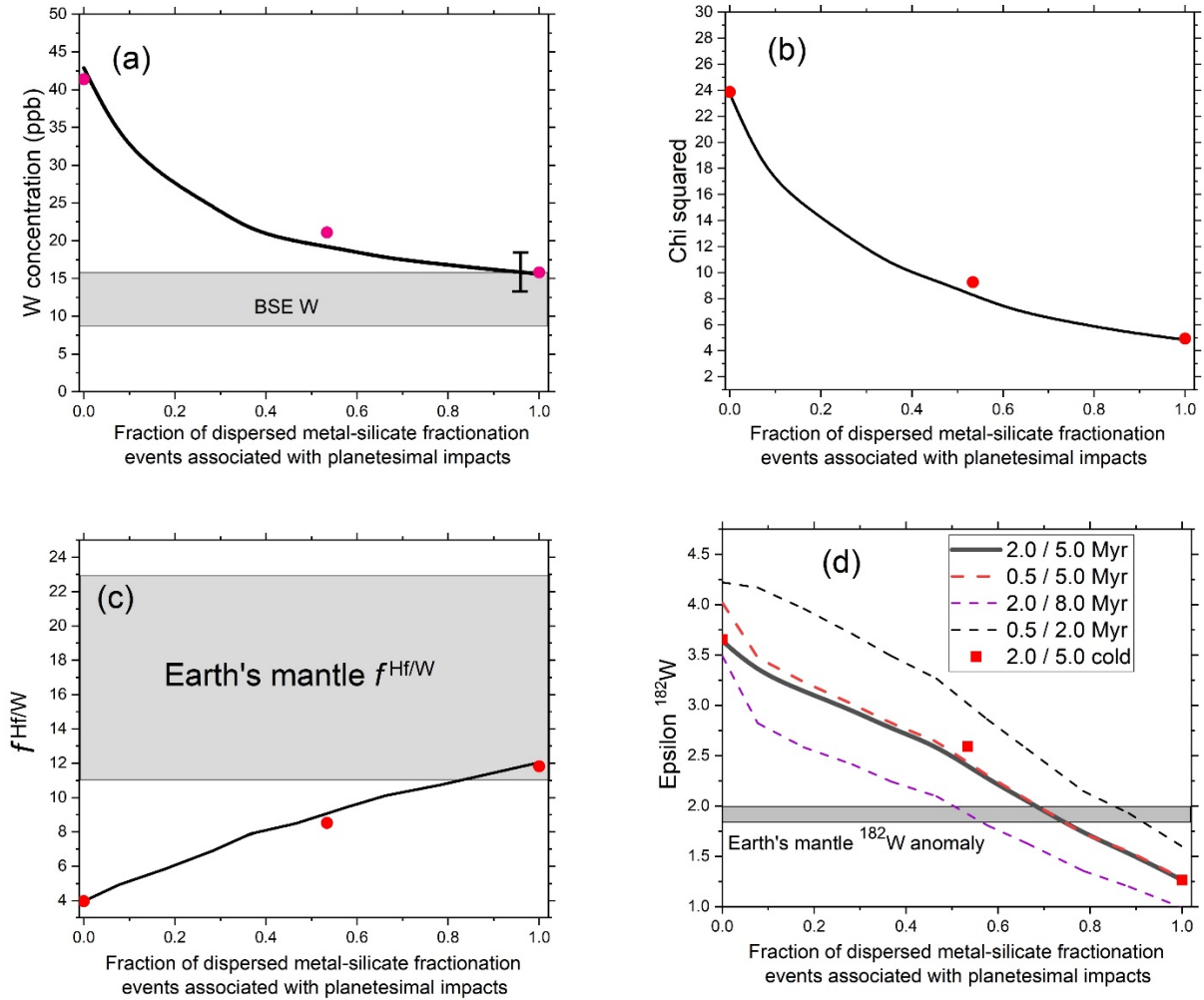

**Fig. 3.** Effects of the fraction of metal-silicate fractionation events that occur by the dispersed mechanism, following planetesimal impacts, on the results of simulation 1 (4:1-0.5-8) accretion/core formation model. (a) Calculated W concentration compared with the BSE value reported by König et al. (2011) and Palme and O'Neill (2014). The error bar is based on propagated uncertainties in the metal-silicate partition coefficient. (b) Reduced chi squared based on calculated MgO, FeO, SiO$_2$, Ni, Co, Nb, Ta, V, Cr, W and Mo concentrations fitted to those of the primitive mantle (Palme and O'Neill, 2014; Greber et al., 2015). (c) The final mantle Hf/W ratio as defined by $f^{Hf/W}$. (d) Final $\varepsilon_{182W}$ values for different chronological scenarios. In all cases, metal-silicate equilibration occurs at the time of each impact ("continuous magma ocean" model – see below). The solid black curve is based on core-mantle differentiation in starting bodies occurring 2 Myr after the start of the solar system and the main stage of accretion (as modeled in the N-body simulation) starting 5 Myr after the start of the solar system. The three

dashed curves show $\varepsilon_{182W}$ results when these two respective values are 0.5/5.0 Myr, 2.0/8.0 Myr, and 0.5/2.0 Myr. The Earth's mantle $f^{Hf/W}$ value shown in (c) is from Nimmo and Kleine (2015) and Kleine and Walker (2017) and the value of $\varepsilon_{182W}$ in (d) is from Kleine et al. (2002, 2009). The main results shown is this figure are obtained using the "warm target" model of giant impact induced melting with an initial target surface temperature of 2000 K (Nakajima et al., 2021); when the thermal state of the target is defined by an initial surface temperature of 300 K (red symbols) the results are almost identical.

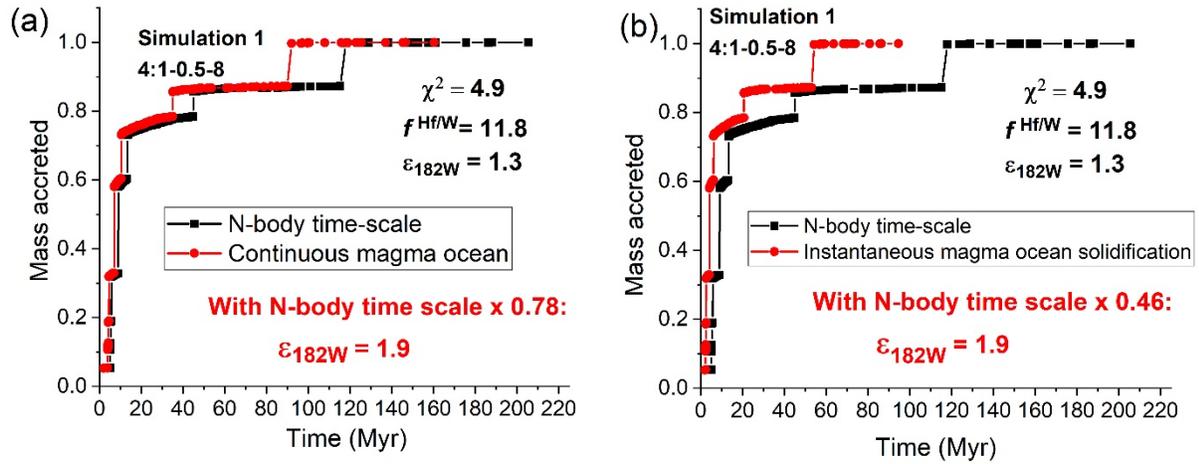

**Fig. 4**. Accretion histories of the Earth-like planet in N-body accretion simulation 1 (4:1-0.5-8) (Rubie et al., 2015). Mass accreted, normalized to one Earth mass, is plotted against time and each symbol represents an accretional event. The black symbols/lines show the original N-body time scale assuming that accretion starts 5 Myr after the start of the solar system, which results in $\varepsilon_{182W} = 1.3$ when accreted metal equilibrates at the time of each impact. The red symbols/lines show the Earth's mantle ($\varepsilon_{182W} = 1.9$) for two extreme core-mantle differentiation models. (a) Results of the "continuous magma ocean model" with the final, moon-forming, giant impact occurring at 92 Myr. (b) Results of the "instantaneous magma ocean solidification model" with the final giant impact occurring at 54 Myr.

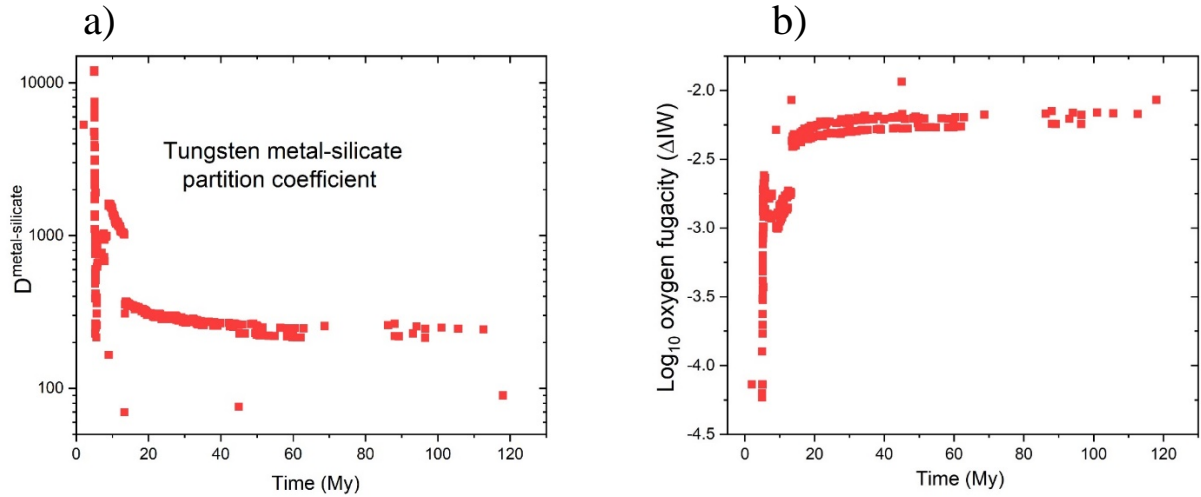

**Fig. 5**. (a) Values of the metal-silicate partition coefficient of tungsten for all metal-silicate equilibration events in accretion simulation 1 (4:1-0.5-8), shown as a function of time after the start of the solar system. (b) $\text{Log}_{10}$ oxygen fugacity values, relative to the iron-wüstite buffer, determined for all metal-silicate equilibration events in the simulation, as a function of time. The fact that the trend is close to being an inverted version of the trend of (a) shows that oxygen fugacity is a dominant factor that controls the metal-silicate partition coefficient of W.

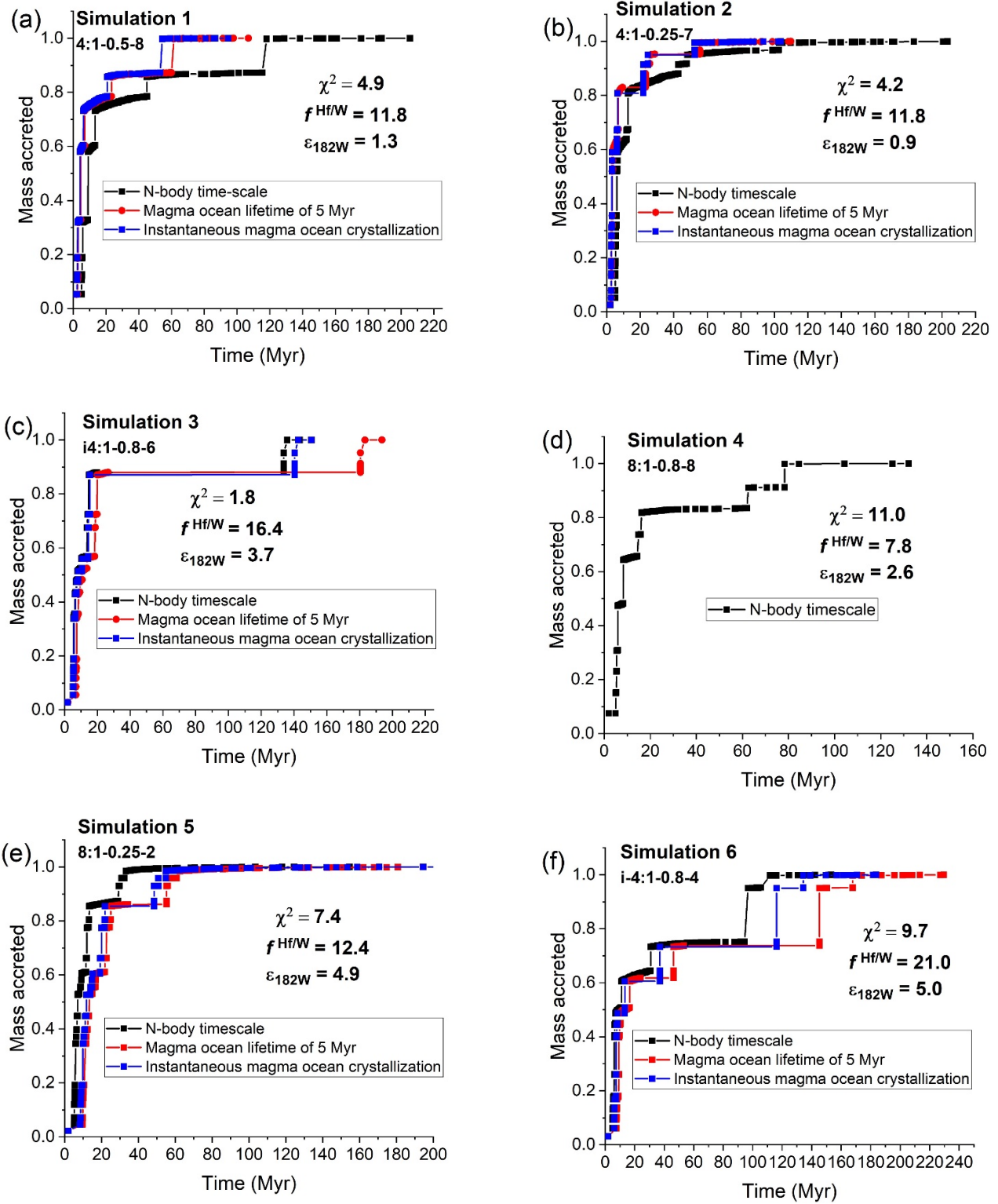

**Fig. 6**. Accretion histories of the Earth-like planets in the six N-body accretion simulations of Rubie et al. (2015) and Jennings et al. (2021). Mass accreted, normalized to one Earth mass, is plotted against time and each symbol represents an accretional event. The black symbols/lines

show the original N-body time scales (without any correction) and the W anomaly listed in black is based on accreted metal equilibrating at the time of each impact. Results are shown for the instantaneous magma ocean crystallization model (blue symbols/lines) and the magma ocean lifetime of 5 Myr model (red symbols/lines) for which the respective timescales have been adjusted by the factors t_adjust (Table 1) that change the tungsten isotope anomaly of Earth's mantle to $\varepsilon_{182W}$ = 1.9. The W isotope anomalies calculated prior to making this artificial timescale adjustment are listed in Table 1. These adjustments have not been made in the case of simulation 4 (d) because it satisfies none of the BSE constraints (Table 1).

# Tungsten isotope evolution during Earth's formation and new constraints on the viability of accretion simulations

## Supplementary Material

**Fig. S1:**

Simulation 2
(4:1-0.25-7)

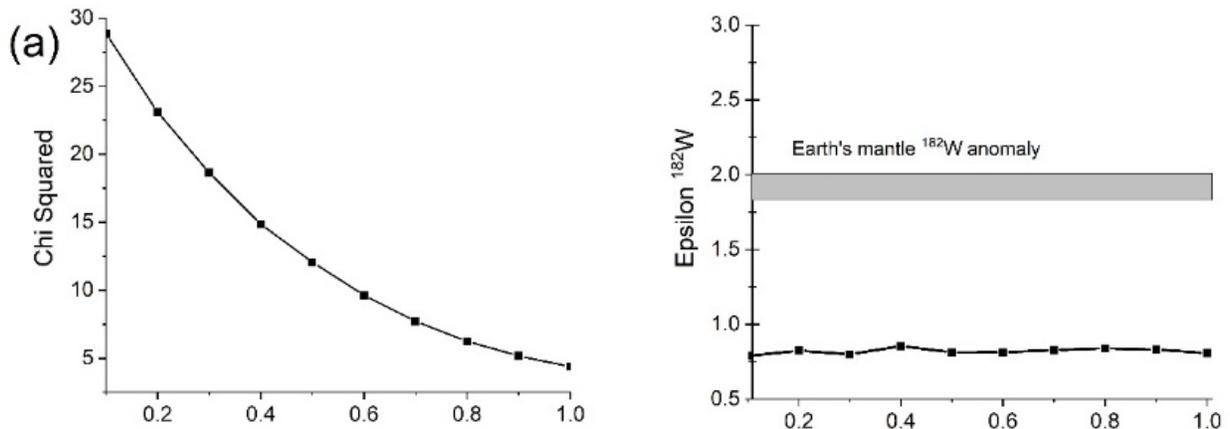

Average fraction of embryo cores that equilibrates with silicate melt

Simulation 3
(i4:1-0.8-6)

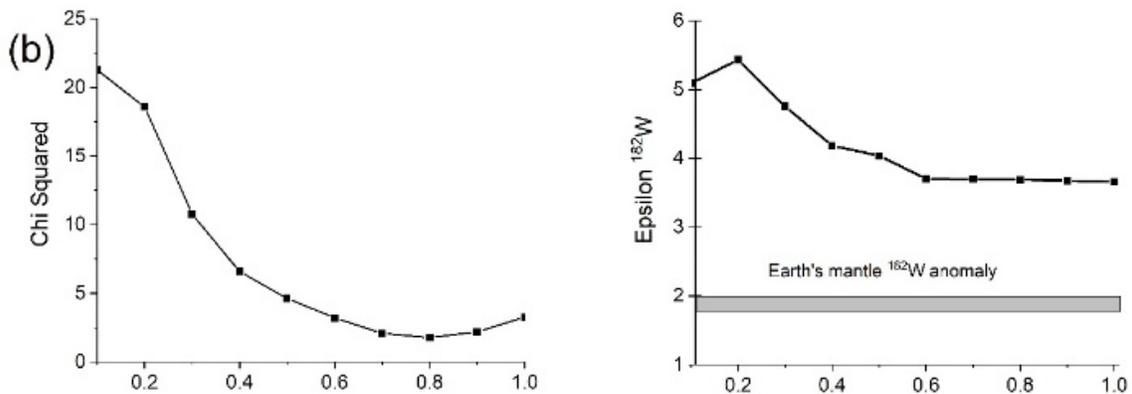

Average fraction of embryo cores that equilibrates with silicate melt

## Simulation 4
## (8:1-0.8-8)

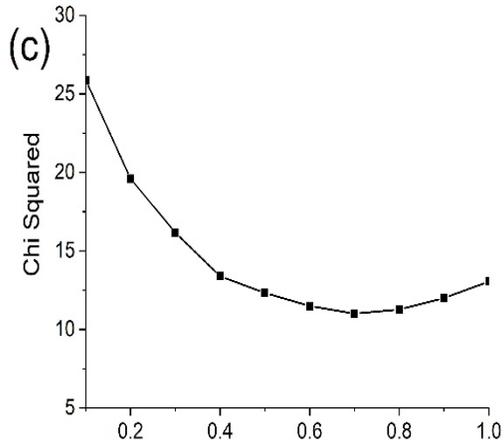
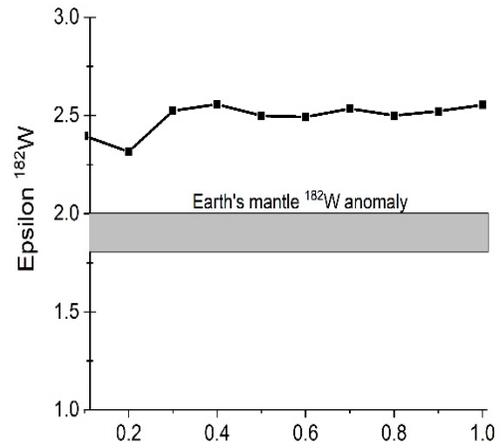

Average fraction of embryo cores that equilibrates with silicate melt

## Simulation 5
## (8:1-0.25-2)

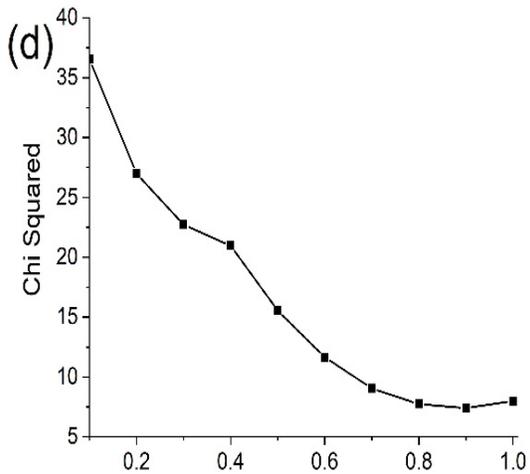
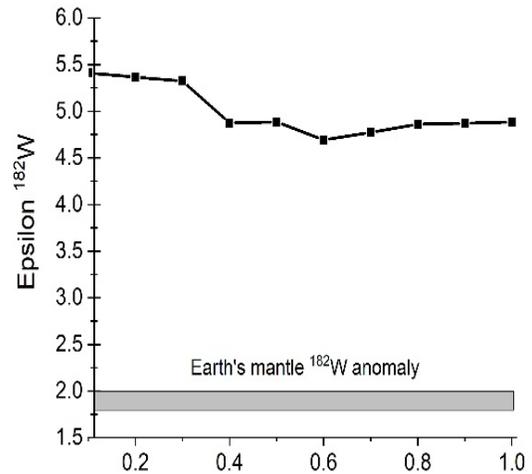

Average fraction of embryo cores that equilibrates with silicate melt

Simulation 6

(i4:1-0.8-4)

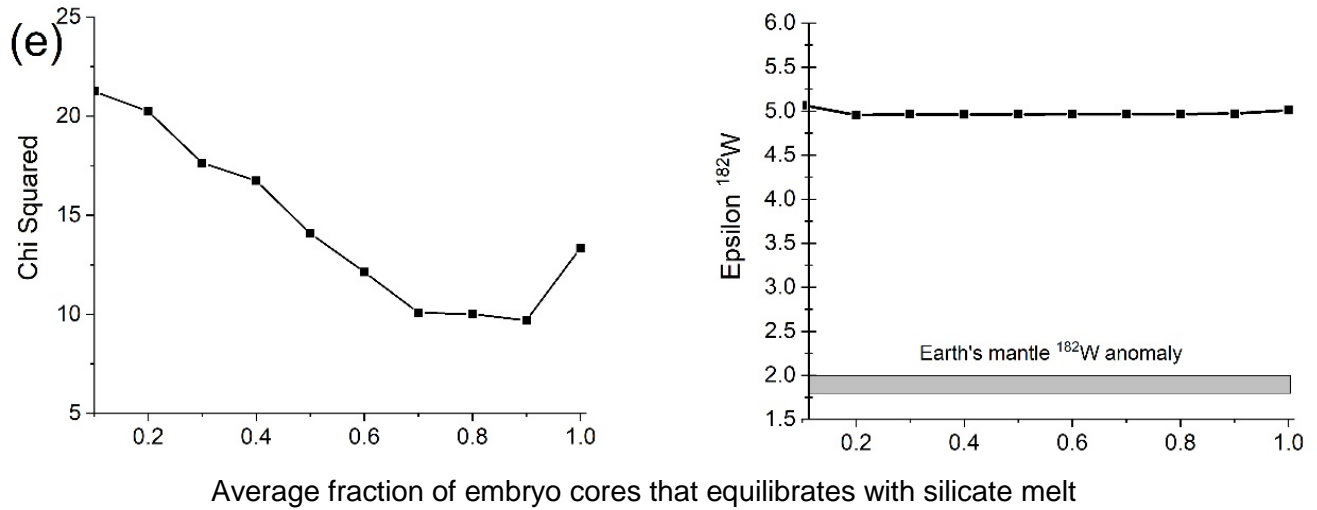

Average fraction of embryo cores that equilibrates with silicate melt

**Fig. S1**. The effects on chi squared and epsilon $^{182}$W of varying the average fraction of embryo cores that equilibrates with silicate liquid at the base of giant impact-induced melt pools in simulations 2-6. Results are shown for fractions of dispersed metal-silicate fractionation events for planetesimal impacts equal to 1.0. Reduced chi squared is determined from fitting calculated mantle concentrations of MgO, FeO, SiO$_2$, Ni, Co, Nb, Ta, V, Cr, W and Mo to BSE values (König et al., 2011; Palme and O'Neill, 2014; Greber et al., 2015). Results for simulation 1 are shown in Fig. 2.

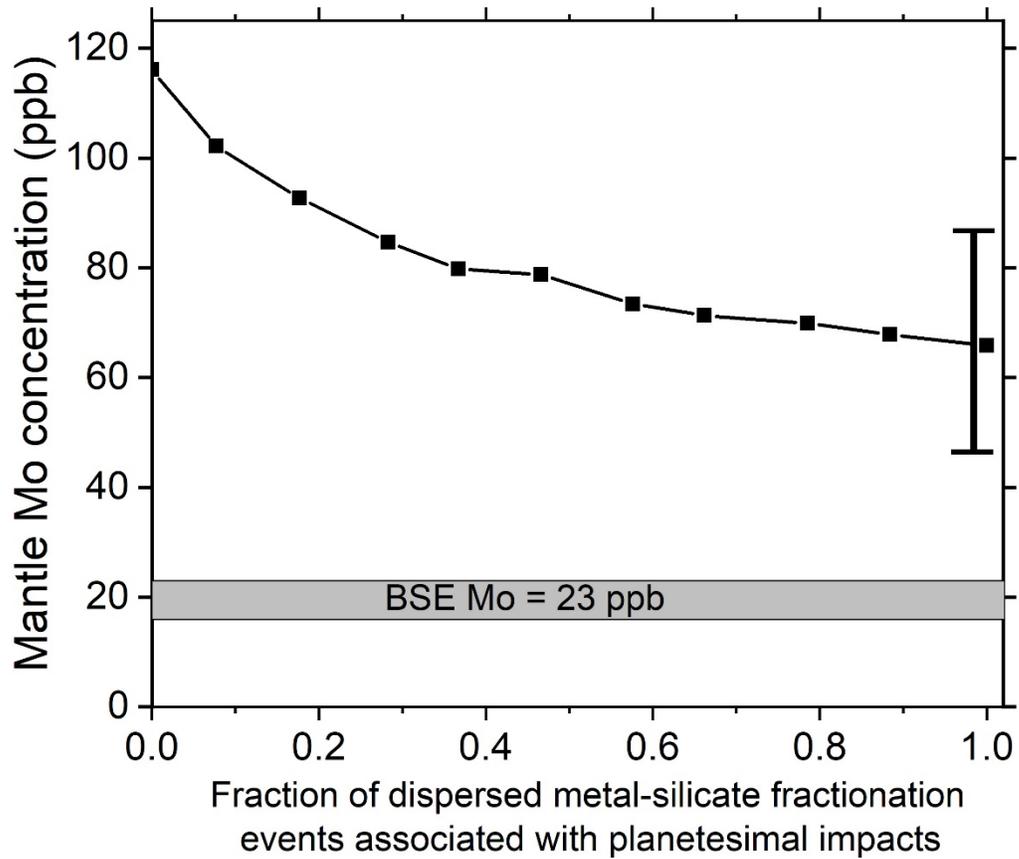

**Fig. S2**. Effects of the fraction of metal-silicate fractionation events that occur by the dispersed mechanism, following planetesimal impacts, on the calculated Mo concentration compared with the BSE value. The accretion simulation is simulation 1 (4:1-0.5-8) as in Fig. 3. The error bar is based on propagated uncertainties in the metal-silicate partition coefficient. Only in simulation 3 is the BSE Mo concentration achieved in the least squares fit (Table 1).

# Calculated compositions of Earth's mantle and core and giant impact history for Grand Tack simulation 3.

(BSE values from König et al. 2011 and Palme & O'Neill 2014; Mo from Grever et al. 2015 and C from Hirschmann 2018)

## Simulation 3 (i4:1-0.8-6)

**FINAL WT% COMPOSITIONS** Components included in chi square calculation indicated by *

| MANTLE | | BSE |
|---|---|---|
| $Al_2O_3$* | 4.55e+00 (5.1e-03) | 4.49e+00 (3.5e-01) |
| MgO* | 3.71+01 (4.1e-02) | 3.67e+01 (3.6e-01) |
| CaO* | 3.65e+00 (4.1e-03) | 3.65e+00 (2.9e-01) |
| Na | 2.48e-01 (2e-04) | 2.59e-01 (1.3e-02) |
| FeO* | 8.10e+00 (7e-03) | 8.10e+00 (5.0e-02) |
| $SiO_2$* | 4.55e+01 (4.9e-02) | 4.54e+01 (3.2e-01) |
| Ni* | 1.67e-01 (2.7e-02) | 1.86e-01 (9.3e-03) |
| Co* | 1.07e-02 (1.7e-03) | 1.02e-02 (5.1e-04) |
| Nb* | 5.75e-05 (1.e-06) | 5.95e-05 (1.2e-05) |
| Ta* | 4.09e-06 (4.7e-08) | 4.30e-06 (2.2e-07) |
| V* | 8.72e-03 (2.0e-03) | 8.60e-03 (4.3e-04) |
| Cr* | 2.92e-01 (3.3e-02) | 2.52e-01 (2.5e-02) |
| W* | 1.168e-06 (5.7e-07) | 1.20e-06 (3.6e-07) |
| Mo* | 4.63e-06 (2.01e-06) | 2.30e-06 (9.0e-07) |
| $H_2O$ | 9.50e-02 | 1.0e-01 (3e-02) |
| C | 1.15e-02 | 1.40e-02 (4.0e-03) |

**Chi Sq = 1.80**

| CORE | |
|---|---|
| Fe | 7.91e+01 (7e-03) |
| Si | 8.13e+00 (5e-02) |
| Ni | 5.05e+00 (3e-02) |
| Co | 2.32e-01 (2e-03) |
| Nb | 4.96e-05 (1.8e-06) |
| Ta | 4.32e-07 (4.8e-08) |
| V | 1.16e-02 (2.0e-03) |
| Cr | 6.83e-01 (3.3e-02) |
| W | 5.56e-05 (6e-07) |
| Mo | 5.71e-04 (2e-06) |
| O | 4.9e+00 |
| H | 8.3e-03 |
| C | 9.06e-02 |

**EARTH'S GIANT IMPACT PARAMETERS:**
**MELT POND, WARM TARGET model (Nakajima et al., 2021)**
**Magma ocean lifetime of 5 Myr model (Table 1, Fig. 6c)**

| Time (yrs)[1] | impact angle[2] | impact vel[3] | Mass ratio[4] | P_eq (GPa)[5] | pf (GI)[6] | Δt(GI) (Myr)[7] |
|---|---|---|---|---|---|---|
| 6801300 | 33.25 | 1.023 | 0.940 | 9.949 | 0.808 | |
| 6820298 | 58.34 | 1.060 | 0.528 | 16.71 | 1.000 | 0.019 |
| 6891779 | 16.77 | 1.013 | 0.338 | 18.20 | 0.866 | 0.072 |
| 6925365 | 29.27 | 1.047 | 0.212 | 21.73 | 0.882 | 0.034 |
| 7186988 | 48.89 | 1.075 | 0.168 | 29.37 | 0.984 | 0.262 |
| 7314333 | 70.25 | 1.056 | 0.781 | 48.66 | 1.000 | 0.127 |
| 8414685 | 15.01 | 1.077 | 0.203 | 6.745 | 0.113 | 1.101 |
| 9353583 | 70.26 | 1.111 | 0.087 | 64.16 | 0.981 | 0.939 |
| 10695469 | 72.13 | 2.094 | 0.064 | 67.98 | 0.966 | 1.342 |
| 13709655 | 62.22 | 1.565 | 0.067 | 73.74 | 0.967 | 3.01 |
| 18461209 | 70.13 | 1.321 | 0.185 | 89.51 | 0.979 | 4.75 |
| 18830488 | 81.04 | 1.232 | 0.070 | 74.65 | 0.762 | 0.37 |
| 19835833 | 69.63 | 1.275 | 0.200 | 100.4 | 0.846 | 1.01 |
| 180526050 | 38.79 | 1.728 | 0.043 | 43.60 | 0.334 | 160.69 |
| 183343500 | 56.49 | 1.069 | 0.050 | 64.78 | 0.471 | 2.814 |

[1] N-body simulation times $\times 1.35$ in order to obtain $\varepsilon_{182W} = 1.9$, as shown in Fig, 6c

[2] Impact angle: 90°=head on impact; 0°=grazing impact;   [3] Impact vel = impact velocity/mutual escape velocity

[4] Mass ratio = mass of impactor/mass of target;   [5] Metal-silicate equilibration pressure

[6] $pf$ = metal-silicate equilibration pressure/CMB pressure at time of impact

[7] $\Delta t(GI)$ = Time since previous giant impact in Myr.

**Average $pf$ for all planetesimal impacts = 0.33**